\documentclass[aps,pra,10pt,twocolumn,longbibliography,noeprint]{revtex4-2}
\usepackage[hidelinks]{hyperref}
\usepackage{graphicx}%
\usepackage{mathtools,amssymb,amsfonts}%

\newcommand{\ket}[1]{|\mathopen#1\rangle}
\newcommand{\I}{\mathbb{I}}
\usepackage{xspace}
\usepackage{xcolor}
\newcommand{\CNOT}{\textsc{cnot}\xspace}
\usepackage{tikz}
\usepackage{comment}

\begin{document}
\title{The Contextual Heisenberg Microscope}

\author{Jan-Åke Larsson}\email{jan-ake.larsson@liu.se}
\affiliation{Department of Electrical Engineering,\\ Link\"oping University,\\ 58183 Link\"oping, SWEDEN}

\begin{abstract}
The Heisenberg microscope provides a powerful mental image of the measurement process of quantum mechanics (QM), attempting to explain the uncertainty relation through an uncontrollable back-action from the measurement device.
However, Heisenberg's proposed back-action uses features that are not present in the QM description of the world, and according to Bohr not present in the world.
Therefore, Bohr argues, the mental image proposed by Heisenberg should be avoided.
Later developments by Bell and Kochen-Specker shows that a model that contains the features used for the Heisenberg microscope is in principle possible but must necessarily be nonlocal and contextual.
In this paper we will re-examine the measurement process within a restriction of QM known as Stabilizer QM (SQM), that still exhibits for example Greenberger-Horne-Zeilinger nonlocality and Peres-Mermin contextuality.
The re-examination will use a recent extension of SQM, the Contextual Ontological Model (COM), where the system state gives a complete description of future measurement outcomes reproducing the quantum predictions, including the mentioned phenomena.
We will see that the resulting contextual Heisenberg microscope back-action can be completely described within COM, and that the associated randomness originates in the initial state of the pointer system, exactly as in the original description of the Heisenberg microscope.
The presence of contextuality, usually seen as prohibiting ontological models, suggests that the contextual Heisenberg microscope picture could perhaps be enabled in general QM.
\end{abstract}

\keywords{Heisenberg microscope, Measurement problem, EPR completeness, Contextuality}

\maketitle

\section{Introduction}%
\begin{figure}[t]
\centering
\includegraphics{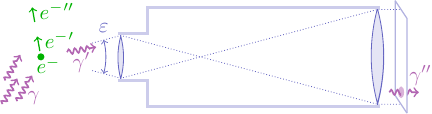}
\caption{The Heisenberg microscope. To measure the position of an electron $e^-$ illuminate it with a pointer $\gamma$ ray, that is scattered to $\gamma'$ causing a recoil in the electron $e^-{}'$ and entangling the two. The scattered $\gamma'$ position can then be measured through microscope optics, say creating a spot on a screen. The post-measurement state of $e^-{}'',\gamma''$ is separable.}%
\label{fig:HM}
\end{figure}

%\label{sec:intro}%
%
The uncertainty relation \cite{Heisenberg1927,Heisenberg1927a} is a fundamental property of Quantum Mechanics, often regarded as the most distinctive feature in which QM differs from classical mechanics.
Heisenberg illustrated the uncertainty relation via a measurement of the position of an electron using a microscope, a setup later known as the Heisenberg microscope (see Fig.~\ref{fig:HM}). 
The accuracy of such a measurement is limited by the illumination wavelength, so to perform an accurate position measurement one would need light of a very short wavelength. 
For such short wavelengths, the interaction with an electron should then be considered as a collision of at least one photon through the Compton effect~\cite{Compton1923}. 
And in such a collision, the electron would recoil, changing its momentum~\cite{Heisenberg1927}. 
Aided by Bohr (note added in proof \cite{Heisenberg1927}), Heisenberg later correctly connected the size of the uncertainty to the ratio between the wavelength and the size of the microscope aperture~\cite{Heisenberg1930}.
The uncertainty relation \cite{Kennard1927} reads
\begin{equation}
\sigma_x\sigma_p\ge\frac \hbar2.
\end{equation}

This first formulation of the uncertainty relation could be read as an epistemological statement, since it allows the possibility that an electron can possess both a well-defined position and momentum but there is a limit to what we can know about the electron.
In that setting, the Heisenberg microscope would indeed output the well-defined position through a collision that induces a change in the well-defined momentum.
However, Heisenberg \cite{Heisenberg1927a} writes in his conclusions that ``such speculations seem to us, to say it explicitly, fruitless and senseless.''
Bohr fills in \cite{Bohr1928}: ``Now the quantum postulate implies that any observation of atomic phenomena will involve an interaction with the agency of observation not to be neglected. Accordingly, an independent reality in the ordinary physical sense can neither be ascribed to the phenomena nor to the agencies of observation.''
Although there is no direct deductive path between a forced inclusion of the agency of observation and the nonexistence of an independent reality of the observed phenomena, Bohr's continued argument over several years established the notion of \textit{complementarity} which is today widely seen as a foundational principle of QM.

Further developments by Bell \cite{Bell1964} and Kochen-Specker \cite{Kochen1967,Budroni2022} (KS) tell us that any possible independent reality must give nonlocal and contextual outcomes from the pre-existing values: future outcomes must sometimes depend on which measurements have been performed, even if the future measurements commute with the performed ones.
These no-go results are usually taken as strong arguments against an independent reality of the observed phenomena, even supported by experiment~\cite{Hensen2015,Giustina2015,Shalm2015}.

\begin{figure}[t]
\centering
\includegraphics{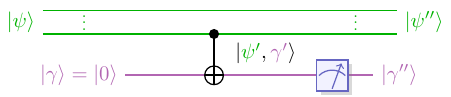}
\caption{The qubit Heisenberg microscope.
To measure in the computational basis (qubit ``position'') on one of many qubits in the joint state $\ket\psi$  let the measured qubit interact with a pointer qubit $\ket\gamma=\ket0$, perhaps of a different type where you have access to a measurement device, through a \CNOT (qubit ``collision'') to create the entangled $\ket{\psi',\gamma'}$, then measure on the pointer qubit.
The post-measurement state $\ket{\psi'',\gamma''}$ is separable.}%
\label{fig:QbitHM}
\end{figure}

In quantum computing the role of position is played by the ``computational basis'' $\ket0$ and $\ket1$, the eigenstates of the $Z$ operator (see Fig.~\ref{fig:QbitHM}), and momentum by the ``phase basis'' $\ket+=(\ket0+\ket1)/{\sqrt2}$ and $\ket-=(\ket0-\ket1)/{\sqrt2}$, the eigenstates of the canonically conjugate $X$ operator \cite{Nielsen2010}.
The differential operator associated with momentum is replaced by difference operators in this discrete case as indicated in the $\ket-$ element of the phase basis. To write down the uncertainty relation we will need a third basis 
$\ket{+i}=(\ket0+i\ket1)/{\sqrt2}$ and $\ket{-i}=(\ket0-i\ket1)/{\sqrt2}$, the eigenstates of the third operator $Y=iXZ$, and we obtain
\begin{equation}
\sigma_Z\sigma_Y\ge\bigl|\langle X\rangle\bigr|.
\end{equation}
A simple way to read this is that if one can predict the outcome of measuring $X$ to be either $+1$ or $-1$ with probability 1, then $\bigl|\langle X\rangle\bigr|=1$ which forces both $\sigma_Z$ and $\sigma_Y$ to reach their maximal values 1. Complete knowledge of $X$ forces complete uncertainty of both $Z$ and $Y$.
The elastic impact of the Compton effect also has a direct correspondence in the \CNOT gate present in Fig.~\ref{fig:QbitHM}.

Qubit quantum computing still contains noncommuting observables and the uncertainty relation even when restricting to only Pauli-tensor-product (Pauli-group) observables and preparations, so that one obtains Stabilizer QM (SQM) \cite{Gottesman1997,Gottesman1998a}, and also displays Bell nonlocality through the Greenberger-Horne-Zeilinger (GHZ) paradox \cite{Greenberger1990} and KS contextuality through the Peres-Mermin square \cite{Peres1990,Mermin1990}.
As previously mentioned, both of these no-go results strengthen the argument that Bohr's quoted conclusion holds, here also for SQM. 

However, SQM can be extended into the Contextual Ontological Model (COM) \cite{Hindlycke2022} that reproduces all predictions of SQM including Bell nonlocality and KS contextuality, even though all measurement outcomes are generated deterministically from the COM state. 
Thus, in contrast to Bohr's claim, {an independent reality in the ordinary physical sense can be ascribed to the phenomena} of SQM. 
This paper intends to show that this also applies to {the agencies of observation}, including the state update of a system measured through the Heisenberg microscope.

We start with a walk-through of Stabilizer QM and the Contextual Ontological Model, needed for the argument, and then showcase the contextual Heisenberg microscope.
A detailed argument can be found in Appendixes~\ref{sec:AppA}--\ref{sec:AppC}.

\section{Stabilizer Quantum Mechanics}
SQM represents QM states not as complex vectors but indirectly as a set of Pauli-group observables that stabilize (preserve) the state.
For example, $\ket0$ is the unique state stabilized by $+Z$, $\ket1$ by $-Z$, $\ket+$ by $+X$, and $\ket-$ by $-X$. 
For more qubits, the available states include tensor products of these but also some entangled states such as $(\ket{00}+\ket{11})/\sqrt2$, the unique state stabilized by both $XX$ and $ZZ$.
The tensor product sign has been omitted here and below as is standard in SQM, so for clarity we write $\times$ for multiplication.
To obtain a unique stabilized state it is enough to write down a small set of commuting stabilizers (the \textit{generators} of the \textit{stabilizer group}, details in Appendix~\ref{sec:AppA}), this is the motivation to use SQM.

SQM only has access to measurement of Pauli-group observables, so there are two alternatives: either the measured observable commutes with all generators, and then the outcome is fixed by the generators, otherwise the outcome is random $\pm1$ with equal probability.
For example, the state $\ket{01}$ is uniquely stabilized by $Z\I$ and $-\I Z$, and the observable $ZZ$ commutes with both.
An explicit calculation gives us $-ZZ=Z\I\times(-\I Z)$ so $-ZZ$ stabilizes $\ket{01}$; the measurement outcome for $ZZ$ will be $-1$.
The QM state is unchanged under this measurement.

Calculating the outcome requires us to find which generators are contained in the observable.
This calculation is much simpler to do when using a helper set of \textit{destabilizers} \cite{Aaronson2004}, one Pauli-group observable per generator that does not commute with that generator but commutes with all the others.
In our example, the generators $Z\I$ and $-\I Z$ have destabilizers $X\I$ and $\I X$; note that $X\I$ does not commute with $Z\I$ but does commute with $-\I Z$, and vice versa for $\I X$.
Given a Pauli-group observable that commutes with all generators, it is now a simple matter of checking which destabilizers do not commute with the observable; their corresponding generators then build the observable.
The $ZZ$ observable does not commute with either $X\I$ or $\I X$ so both $ZI$ and $-IZ$ need to be used to build the observable, as we already saw.

Adding the helper set of destabilizers gives us the Stabilizer State Tableau Representation (SSTR) \cite{Aaronson2004}. The SSTR state consists of a set of stabilizer generators paired with equally many destabilizers, for example
\begin{equation}
	\ket\psi=(\ket{00}+\ket{11})/\sqrt2\simeq\{\underset{\mathclap{\text{stab}}}{XX}\tikz{\useasboundingbox; \draw[<->,bend right=10,semithick](0,-.15)to+(.85,0);}
  ,\overset{\mathclap{\text{stab}}}{ZZ}\tikz{\useasboundingbox;\draw[<->,bend left=10,semithick](.05,.4)to+(.7,0);}
  ;\;\underset{\mathclap{\text{\quad\ destab}}}{Z\I},\overset{\mathclap{\text{\quad\ destab}}}{\I X}\}.
  \label{eq:SSTRpsi}
\end{equation}
Here the stabilizer generators $XX$, $ZZ$ commute, the destabilizer $Z\I$ does not commute with $XX$ but does commute with $ZZ$, and similarly the destabilizer $\I X$ does not commute with $ZZ$ but does commute with $XX$.

\begin{figure}[t]
	\centering
	\includegraphics{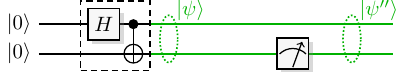}
	\caption{%
  Pauli-$Z$ measurement on the last qubit of the entangled  $\ket\psi=(\ket{00}+\ket{11})/{\sqrt2}$, stabilized by $XX$ and $ZZ$. 
	}
	\label{fig:SSTR}
\end{figure}

QM tells us that a Pauli-group observable that does not commute with all stabilizers will have equally probable measurement outcomes $\pm1$ \cite{Aaronson2004}, often attributed to ``true'' or ``quantum'' randomness.
The QM state changes under such a measurement, so the stabilizers need to be updated.
As an example see Fig.~\ref{fig:SSTR}. Measuring the observable $\I Z$ on the state in Eq.~(\ref{eq:SSTRpsi}), we find that $\I Z$ does not commute with $XX$, so the outcome is
\begin{equation}
v(\I Z)=a,\quad a=\pm1\text{ equally probable.}
\end{equation}
The post-measurement state $\ket{\psi''}$ is $\ket{00}$ if $a=+1$ and $\ket{11}$ if $a=-1$, both stabilized by $a\I Z$ and $ZZ$.
The previous stabilizer $XX$ that didn't commute with the observable $\I Z$ is replaced by $a\I Z$, and $ZZ$ is preserved. 
The replaced $XX$ is instead used as a destabilizer in the new SSTR state.
The final step is to make the remaining destabilizer $\I X$ commute with the new stabilizer $a\I Z$, this can be achieved by replacing $\I X$ with $\I X\times XX=X\I$ which does commute with $a\I Z$.
The new SQM state is
\begin{equation}
\begin{split}
\ket{\psi''}&=  \begin{cases}
  \ket{00}\text{ if }a=+1,\\\ket{11}\text{ if }a=-1\\
  \end{cases}\\
&\simeq \{a\I Z,ZZ;\;XX,X\I\}.
\end{split}
\end{equation}
The general update mechanism is described in Appendix~\ref{sec:AppA}.

\section{The Contextual Ontological Model}
The key insight of COM \cite{Hindlycke2022} is that the set of stabilizers and destabilizers form a basis for the Pauli group, so \textit{any} Pauli-group observable can be expressed in the elements, perhaps multiplied with an additional $\pm1$ phase.
For example, the observable $\I Z=ZZ\times Z\I$, using one stabilizer and one destabilizer from the state in Eq.~(\ref{eq:SSTRpsi}).

In COM the additional $\pm1$ phase is used as the measurement outcome for {any} Pauli-group observable~\cite{Hindlycke2022}. 
Thus, outcomes of all available measurements are predicted with probability 1, so COM is Einstein-Podolsky-Rosen (EPR) complete~\cite{Einstein1935}.
To reproduce SQM predictions for any single Pauli measurement, it is enough for COM to initialize the destabilizer phases to be random $\pm1$ equally probable and independent. 
The COM state $\ket\psi$ in Fig.~\ref{fig:SSTR} would be
\begin{equation}
\begin{split}
	\ket\psi=(\ket{00}+\ket{11})/\sqrt2\simeq\{XX,ZZ;\;aZ\I,b\I X\},\\
  a,b=\pm1\text{ eq prob indep.}
  \label{eq:COMpsi}
\end{split}
\end{equation}
From this state we obtain $a\I Z=ZZ\times aZ\I$, so the COM state in Eq.~(\ref{eq:COMpsi}) fixes the $\I Z$ measurement outcome 
\begin{equation}
v(\I Z)=a.
\end{equation}
This measurement outcome is fixed given the COM state, note that this is completely different from SSTR where the destabilizer phases are never used and instead a random outcome is generated on demand at measurement.

The measurement state-update of COM is similar to that of SQM but has another crucial difference: an additional last step to randomize the phase of the destabilizer paired with the measured observable.
This last step occurs even if the observable commutes with all generators, in which case the SSTR measurement update did nothing.

In our example the post-measurement state still is $\ket{00}$ if $a=+1$ and $\ket{11}$ if $a=-1$, both stabilized by $a\I Z$ (that replaces the noncommuting $XX$) and $ZZ$.
The replaced $XX$ is used as a destabilizer just as in SSTR, and the remaining destabilizer $b\I X$ is replaced by $b\I X\times XX=bX\I$ using the same mechanism as SQM to preserve the commutation relations. 
The crucial difference to SQM is a new final step, to randomize the phase of the new destabilizer so that it becomes $cXX$, creating the post-measurement state
\begin{equation}
\begin{split}
  \ket{\psi''}&=
  \begin{cases}
  \ket{00}\text{ if }a=+1\\\ket{11}\text{ if }a=-1\\
  \end{cases}\\
  &\simeq \{a\I Z,ZZ;\;cXX,bX\I\},\\
 &\hspace{15mm} c=\pm1\text{ eq prob.}
\label{eq:COMpsibis}
\end{split}
\end{equation}
One could say the quantum randomness is applied after the measurement in COM, rather than before as in SQM.
With this random update COM gives identical predictions to those obtained from SQM for any sequence of measurements, including the uncertainty relation, GHZ nonlocality and KS contextuality \cite{Hindlycke2022}.
For the general update mechanism, see Appendix~\ref{sec:AppB}.
From this point on, our aim is to find the origin of the supposedly random phase $c$ present in the post-measurement COM state.

\begin{figure}[t]
	\centering
  \includegraphics{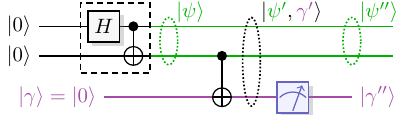}
	\caption{The contextual Heisenberg microscope for a   Pauli-$Z$ measurement on the last qubit of the entangled  $\ket\psi=(\ket{00}+\ket{11})/{\sqrt2}$, stabilized by $XX$ and $ZZ$.}
	\label{fig:CHM}
\end{figure}

\section{The contextual Heisenberg microscope}
It is now possible to use COM to describe the qubit Heisenberg microscope.
In particular it is possible to ask where and how the randomness enters, and we can follow ``the finite and uncontrollable interaction between the objects and the measuring instruments'' \cite{Bohr1935} in detail.
In the example in Fig.~\ref{fig:CHM}, the measured system is in the COM state of Eq.~(\ref{eq:COMpsi}) and the pointer system is in the state
\begin{equation}
\ket{\gamma}=\ket{0}\simeq\{Z;\;cX\},\quad c=\pm1\text{ eq prob}.
\label{eq:COMgamma}
\end{equation}
The reason for the choice of the label name $c$ will become evident below. 
This makes the joint state
\begin{equation}
\begin{split}
\ket{\psi,\gamma}&=(\ket{000}+\ket{110})/\sqrt2\\
&\simeq\{XX\I,ZZ\I,\I\I Z;\;aZ\I\I,b\I X\I,c\I\I X\},\\
&\hspace{19mm} a,b,c=\pm1\text{ eq prob indep.}
\end{split}
\end{equation}
The next step is to entangle the pointer to the measured system through a \CNOT.
The usual \CNOT map $\ket{10}\leftrightarrow\ket{11}$ and $\ket{+-}\leftrightarrow\ket{--}$ that preserves $\ket{00}$, $\ket{01}$, $\ket{++}$ and $\ket{-+}$, corresponds to a mapping between Pauli-group elements $X\I\leftrightarrow XX$ and $\I Z\leftrightarrow ZZ$ that preserves $\I X$ and~$Z\I$, see Appendix~\ref{sec:AppB}.
Here, the \CNOT creates the tripartite entangled GHZ state \cite{Greenberger1990}
\begin{equation}
\begin{split}
\ket{\psi',\gamma'}&=(\ket{000}+\ket{111})/\sqrt2\\
&\simeq\{XXX,ZZ\I,\I ZZ;\;aZ\I\I,b\I XX,c\I\I X\}.
\end{split}
\end{equation}
From this COM state we obtain $a\I\I Z=ZZ\I\times \I ZZ\times aZ\I\I$,
so the same fixed measurement outcome as before,
\begin{equation}
v(\I\I Z)=a.
\end{equation}
The post-measurement state is $\ket{000}$ if $a=+1$ and $\ket{111}$ if $a=-1$, both stabilized by $a\I\I Z$ (that replaces the non-commuting $XXX$), $ZZ\I$ and $\I ZZ$. The replaced $XXX$ is used as a destabilizer, and the remaining destabilizers $b\I XX$ and $c\I\I X$ are replaced by $b\I XX\times XXX=bX\I\I$ and $c\I\I X\times XXX=cXX\I$, respectively. 
The final step is to randomize the phase of the new destabilizer so that it becomes $dXXX$.
The post-measurement COM state~is 
\begin{equation}
\begin{split}
\hspace{-1ex}\ket{\psi'',\gamma''}&=
  \begin{cases}
  \ket{000}\text{ if }a=+1,\\\ket{111}\text{ if }a=-1\\
  \end{cases}\\
&\simeq\{a\I\I Z,ZZ\I,\I ZZ;\;dXXX,bX\I\I,cXX\I\},\\
&\hspace{41mm} d=\pm1\text{ eq prob}.
\end{split}
\end{equation}
This COM state is separable (because the QM state is); replace the last stabilizer $\I ZZ$ with $\I ZZ\times a\I\I Z=a\I Z\I$ and the first destabilizer $dXXX$ with $dXXX\times cXX\I=cd\I\I X$. 
These replacements change the generators but preserve the generated stabilizer group, and the destabilizer relations including in fact all outcome predictions~\cite{Hindlycke2022}, and creates the equivalent COM state
\begin{equation}
\begin{split}
\hspace{-1mm}\ket{\psi'',\gamma''}&=
  \begin{cases}
  \ket{000}\text{ if }a=+1,\\\ket{111}\text{ if }a=-1\\
  \end{cases}\\
&\simeq\{a\I\I Z,ZZ\I,a\I Z\I;\;cd\I\I X,bX\I\I,cXX\I\},\\
&\hspace{40mm} d=\pm1\text{ eq prob}.
\end{split}
\end{equation}
The stabilizers and destabilizers can be separated into two separate systems and the generator order can be freely chosen, so we obtain 
\begin{equation}
\begin{split}
\ket{\psi''}&=
  \begin{cases}
  \ket{00}\text{ if }a=+1,\\\ket{11}\text{ if }a=-1\\
  \end{cases}\\
&\simeq\{a\I Z,ZZ;\;cXX,bX\I\}\\
\ket{\gamma''}&=  \begin{cases}
  \ket{0}\text{ if }a=+1,\\\ket{1}\text{ if }a=-1\\
  \end{cases}\\
&\simeq\{aZ;\;cdX\}\\
&\hspace{14mm} d=\pm1\text{ eq prob}.
\end{split}
\end{equation}
The post-measurement system state $\ket{\psi''}$ is the same as that of Eq~(\ref{eq:COMpsibis}) with the crucial difference that the phase of the destabilizer $cXX$ is not random from the measurement update; it originates from the unknown initial pointer-system destabilizer $cX$ present in Eq.~(\ref{eq:COMgamma}).

We conclude that the state update of the measured system in the contextual Heisenberg microscope is not random, it is pre-determined by the pointer state, just as in the original Heisenberg microscope.
A general proof that this conclusion always holds in the COM description of SQM can be found in Appendix~\ref{sec:AppC}.

\section{Conclusions}
%\label{sec:discussion}%
%
The Contextual Ontological Model (COM) \cite{Hindlycke2022} is an Einstein-Podolsky-Rosen (EPR) complete \cite{Einstein1935} model for Stabilizer Quantum Mechanics (SQM) \cite{Gottesman1997,Gottesman1998,Aaronson2004} that allows simultaneous well-defined preexisting values associated with incompatible measurements.
COM has a random state update at measurement that creates an upper limit to an observer's knowledge about future outcomes of these incompatible measurements, thus reproducing the uncertainty relation. 

In this paper we have shown that within that model, we can restore the role of the Heisenberg microscope to the one first proposed by Heisenberg \cite{Heisenberg1927}, noting that the microscope must be contextual to reproduce the SQM predictions.
The contextual Heisenberg microscope performs a state update that is no longer random, but instead pre-determined by the initial state of the pointer system.
There is no such mechanism in SQM itself and indeed, if one were to follow Bohr's argument, no such mechanism could exist in SQM. 
Nonetheless, the contextual Heisenberg microscope within COM realizes that mechanism, even though COM reproduces all predictions of SQM.
This also shows that the traditional quantum measurement treatment involving a macroscopic measurement device or the environment is not cruical for tracing the origin of the random element causing the uncertainty. 
It suffices to consider a pointer system of the same type as the one measured upon, in our example a qubit system.

Lack of knowledge of the initial state of the pointer system would serve as an epistemic restriction on the initial COM state preparation, e.g., fixing the pointer computational basis state $\ket0$ but leaving a random initial phase of the COM state $\{Z;cX\}$. 
The completely deterministic state update of the contextual Heisenberg microscope will then cause the random initial pointer phase to randomize the existing predicted measurement outcomes of observables that do not commute with the measured observable.
This forms an epistemic horizon from a deterministic law \cite{Fankhauser2025}; even though an EPR complete model can describe the phenomena, an observer cannot obtain complete knowledge about the model state, because the initial state of the pointer is unknown to the observer.

Thus we can conclude, paraphrasing Bohr, that even though any observation of SQM phenomena will involve an interaction with the agency of observation not to be neglected, it remains possible to ascribe an independent reality in the ordinary physical sense to the phenomena as well as to the agencies of observation. 
COM reproduces all the SQM phenomena that are usually taken to forbid an ontological underlying model, including contextuality and nonlocality, and this suggests that the contextual Heisenberg microscope could be useful outside the restriction to SQM

\section{Acknowledgments}
This work is supported by the Swedish Research Council project no 2023-05031.

%\end{document}

%\end{comment}
\appendix

\noindent
\section{Stabilizer Quantum Mechanics}
\label{sec:AppA}%
Stabilizer Quantum Mechanics (SQM) \cite{Gottesman1997,Gottesman1998,Aaronson2004} is a restriction of many-qubit QM in which the QM state is indirectly represented as a set of observables that stabilize (preserve) the QM state.
SQM restricts to the observables of the Pauli group,
\begin{equation}
M=\pm\mathop\bigotimes_{k=1}^n i^{x_kz_k}X^{x_k}\times Z^{z_k},
\label{eq:XZ}
\end{equation}
where $Y=iX\times Z$ and $x_k$ and $z_k$ are bits, in total $2n+1$ bits.
Tensor product signs will be omitted  below as is standard in SQM, and for clarity we write $\times$ for multiplication.
A set of such elements generate a group, and the set of generators is called independent if removing any single generator reduces the group size.
For $n$ qubits, $n$ commuting independent generators $M_k$ generate a group that stabilizes a unique QM state.

A unitary $U$ will act on a stabilizer $M$ through conjugation, because if $M$ stabilizes $\ket\psi$ then $\varphi_U(M)=U\times M\times U^\dagger$ stabilizes $U\ket\psi$.
The unitaries that preserve the Pauli group constitute the Clifford group, generated by the Hadamard $H$, the $\pi/2$ phase rotation $S$, and \CNOT, where e.g.,
\begin{equation}
\begin{split}
\varphi_H(X)=Z,\quad \varphi_S(&X)=Y,\quad \varphi_S(Z)=Z,\\
\varphi_\CNOT(X\I)=XX,&\quad
\varphi_\CNOT(\I Z)=ZZ,\\
\varphi_\CNOT(Z\I)=Z\I,&\quad
\varphi_\CNOT(\I X)=\I X.
\end{split}
\end{equation}

SQM assumes that we have access to sources of qubits in the $\ket0$ state; all other stabilizer states can be generated by using Clifford unitaries.
The available measurements are Pauli-group observables. 
Even though not all Bell measurements \cite{Bell1964} are available, SQM still contains nonlocal phenomena such as the GHZ paradox \cite{Greenberger1990}, and KS contextuality through the Peres-Mermin square~\cite{Mermin1990,Budroni2022}.
In SQM, if an observable $M$ commutes with all stabilizer generators $M_k$, then  
\begin{equation}
M=v(M)\prod_{k=1}^n M_k^{m_k},
\label{eq:SSTR}
\end{equation}
where $v(M)=\pm1$ corresponds to the measurement outcome of $M$ on the stabilized state. 
The $m_k$ exponents are binary, easiest to calculate using a set of {destabilizers} as in %the Stabilizer State Tableau Representation \cite{Aaronson2004} (SSTR).
SSTR \cite{Aaronson2004}.
A destabilizer $C_k$ % associated with a specific $M_k$ 
is a Pauli-group observable %Hermitian Pauli-group element 
that anticommutes with $M_k$ and commutes with the other $M_j$ (and $C_j$), $j\neq k$ , yielding the tableau
\begin{equation}
\begin{array}{rcl}
\{M_1,&\ldots,&M_n;\\C_1,&\ldots,&C_n\},
\end{array}
\end{equation}
that uses in total $2n(2n+1)$ bits.
It is now simple to calculate the exponents $m_k=M\cdot C_k$ (mod 2) using the symplectic product
\begin{equation}
M\cdot M'=\sum_{k=1}^n(x_kz'_k-z_kx'_k),
\label{eq:symplectic}
\end{equation}
which is $0$ (mod 2) if $M$ commutes with $M'$ and otherwise $1$ (mod 2), in which case $M$ anticommutes with $M'$. 
Thus, checking commutation has linear complexity in $n$, and calculating $v(M)$ has quadratic complexity.

After a measurement of an observable $M$ that commutes with all $M_k$ no state update is needed, since the QM state remains unchanged.
If $M$ anticommutes with at least one $M_k$ (we assume $k=1$ without loss of generality; the generator order is freely chosen), the outcome $v(M)=\pm 1$ is randomly drawn with equal probability, and then the $M_k$ need to be updated to contain $v(M)M$ while preserving the part of the stabilizer group that commutes with $M$.
SSTR first uses $M_1$ to rewrite $M_k$ and $C_k$, $k\ge2$, as
\begin{equation}
M_k'=M_1^{c_k}\times M_k;\quad C_k'=M_1^{m_k}\times C_k,
\end{equation}
where the additional exponent $c_k=M\cdot M_k$ (mod 2) is $0$ if $M$ commutes with $M_k$ and otherwise $1$.
This changes the generators but not the stabilizer group, and $M$ now commutes with $M_k'$ and $C_k'$ for $k\ge2$, because if $M$ anticommutes with both $M_1$ and $A$ then $M$ commutes with~$M_1\times A$.
SSTR then sets $C_1'=M_1$ and $M_1'=v(M)M$ giving the new state
\begin{equation}
\begin{array}{rccl}
\{v(M)M,&M_2',&\ldots,&M_n';\\C_1',&C_2',&\ldots,&C_n'\}.
\end{array}
\label{eq:SSTRout}
\end{equation}
The new $M_k'$ uniquely stabilize the post-measurement quantum state, and the new $C_k'$ are destabilizers, making SSTR a faithful representation of SQM~\cite{Aaronson2004}.

\section{The Contextual Ontological Model}
\label{sec:AppB}%
The key insight \cite{Hindlycke2022} of the Contextual Ontological Model (COM) is that the joint set of $M_k$ and $C_k$ constitutes a symplectic basis for the Pauli group.
Because of this, an observable can be expanded as a product of one element of the stabilizer group and one element from the destabilizer group
\begin{equation}
M=v(M)i^w\prod_{k=1}^nM_k^{m_k}\times\prod_{k=1}^n C_k^{c_k},
\label{eq:M}
\end{equation}
using $m_k$ and $c_k$ as before, and $w=0$ or $1$ depending on whether these two elements commute.
Equation (\ref{eq:M}) determines $v(M)=\pm1$; in COM the measurement outcome is taken to equal $v(M)$, regardless of commutation with the stabilizer group.
Therefore, given the COM state, every Pauli-group measurement outcome can be predicted with probability 1, so COM is EPR complete~\cite{Einstein1935}.
COM uses a specific initial distribution of its state, and a slightly modified measurement update with respect to SSTR, and is then able to reproduce all predictions of SQM.
The details are as follows.

In COM a state change occurs at every measurement, even when $M$ commutes with all $M_k$, using the following extension of the SSTR measurement update.
If $M$ does not commute with one or more $M_k$, COM performs the same update as SSTR but without the randomization since $v(M)$ is predetermined.
This is where the contextual behavior arises; note that no randomization is involved in the contextuality, e.g., in Fig.~\ref{fig:SSTR} measurement of $\I Z$ changes the predicted outcomes from measuring $Y\I$: immediately before the measurement $v(Y\I)Y\I=iXX\times (aZ\I\times b\I X)=abY\I$ while after the measurement $v(Y\I)Y\I=i(a\I Z\times ZZ)\times bX\I=-abY\I$.

If $M$ commutes with all $M_k$ it will anticommute with at least one $C_k$ ($k=1$ can be assumed), and then COM preserves $M_k$ and uses $C_1$ to rewrite the $C_k$, $k\ge 2$ as
\begin{equation}
  M_k'=M_k;\quad C_k'=C_1^{m_k}\times C_k.
\end{equation}
This does not change the destabilizer group, but the observable $M$ now commutes with all $M_k'$ and $C_k'$ for $k\ge2$.
COM then sets $C_1'=C_1$ and $M_1'=v(M)M$.
At this point in the state update, both alternatives give a new state on the form of Eq.~\eqref{eq:SSTRout}. 

The final step of the COM state update, in both cases, is to randomize the phase of the possibly new $C_1'$ to $c=\pm1$ with equal probability, and obtain
\begin{equation}
\begin{array}{rccl}
\{v(M)M,&M_2',&\ldots,&M_n';\\cC_1',&C_2',&\ldots,&C_n'\}.
\label{eq:out}
\end{array}
\end{equation}
This final randomization ensures that the uncertainty relation of SQM holds, so that subsequent measurements that do not commute with $M$ will give random outcomes. 

The initial state distribution is chosen as follows: for qubit $1$, the SSTR source state $\ket0$ is stabilized by $M_1=Z\I\ldots\I$ and destabilized by $C_1=a X\I\ldots\I$, where $a=\pm1$ with equal probability, and similar independent and identically distributed for the other qubits.
The COM state initialization is equivalent to measuring Pauli-$Z$ on the qubit and on outcome $-1$ resetting the qubit to $\ket0$ by applying an $X$ gate; thus preparation is measurement just as in regular QM.
Clifford transformations act on Pauli-group elements identically to SSTR.
With these choices the predictions of COM are now identical to those of SQM, and the algorithmic complexity remains quadratic~\cite{Hindlycke2022}.

We again stress that given the COM state, all measurement outcomes are predetermined. 
There are no probabilities present in the COM state (except $p=1$ and $p=0$), but at the same time, the QM state is implicitly present in the description as the unique state stabilized by the $M_k$.
Less obvious is that this is not Bohmian mechanics \cite{Bohm1951} for qubits; the main reason is that Bohmian mechanics uses position as the preferred phase-space coordinate while COM changes the preferred coordinate according to the preceding sequence of measurements~\cite{Skott2024}.
COM essentially keeps track of the QM state preparation rather than the QM state itself.

\section{The contextual Heisenberg microscope}
\label{sec:AppC}%
COM allows us to closely study the measurement process through the qubit Heisenberg microscope in Fig.~\ref{fig:QbitHM} of the main text.
We can transform measurement of any Pauli-group observable on an $n$-qubit system into measurement of $M=\I\ldots\I Z$, i.e., Pauli-$Z$ on the last qubit~\cite{Aaronson2004,Johansson2023}.
The pointer system can be appended at index $n+1$, by appending one $\I$ to the observable $M\to M\I$ and to each generator within the state, also adding one stabilizer and one destabilizer initialized just like the other systems in our model.
This gives the COM state
\begin{equation}
\begin{array}{rccl}
\{M_1\I,&\ldots,&M_n\I,&\I\ldots\I Z;\\
C_1\I,&\ldots,&C_n\I,&c\I\ldots\I X\},
\label{eq:in}
\end{array}
\end{equation}
where $c=\pm1$ with equal probability.

The Heisenberg microscope (see Fig.~\ref{fig:QbitHM}) proceeds to entangle the system and pointer using a \CNOT, and then measure $M'=\I\ldots\I Z$, i.e., Pauli-$Z$ on the pointer system. 
The \CNOT will change the basis elements $M_k\I$ and $C_k\I$ that anticommute with $M\I$ into $M_kX$ and $C_kX$, resp.~(e.g., $\I\ldots\I X\I\rightarrow\I\ldots\I XX$), and leave  $M_k\I$ and $C_k\I$ that commute with $M$ unchanged.
The \CNOT will also change $\I\ldots\I Z$ into $\I\ldots\I ZZ$ and leave $c\I\ldots\I X$ unchanged.
The COM state after the \CNOT then becomes
\begin{equation}
\begin{array}{rccl}
\{M_1X^{c_1},&\ldots,&M_nX^{c_n},&\I\ldots\I ZZ;\\
C_1X^{m_1},&\ldots,&C_nX^{m_n},&c\I\ldots\I X\},
\end{array}
\label{eq:vN}
\end{equation}
where $c_k=M\cdot M_k$ (mod 2) and $m_k=M\cdot C_k$ (mod 2) as before. 
In this state $M\I$ and $M'$ anticommute with the same basis elements, and $M$ anticommutes with at least one $M_k$ or $C_k$, so at least one $c_k$ or $m_k$ equals 1 ($k=1$ can be assumed).
When measuring $M'$ the COM measurement update then outputs
\begin{equation}
\begin{array}{rcccl}
\{v(M')M',&M_2'\I,&\ldots,&M_n'\I,&\I\ldots\I ZZ;\\
dC_1'X,&C_2'\I,&\ldots,&C_n'\I,&cC_1'\I\},
\end{array}
\label{eq:vNstate}
\end{equation}
with $d=\pm1$ with equal probability.
The destabilizer $C_1'X$ is the only one that anticommutes with $v(M')M'$; the state update makes the rightmost destabilizer $C_1'X\times c\I\ldots\I X=cC_1'\I$.

The stabilizer group remains the same if we exchange the first stabilizer $v(M')M'$ for $v(M')M'\times \I\ldots\I ZZ=v(M)M\I$, likewise the destabilizer group remains the same if we exchange the last destabilizer $dC_1'X$ for $d C_1'X\times cC_1'\I=cd\I\ldots\I X$.
This also maintains a symplectic basis because $v(M)M\I$ and $cd\I\ldots\I X$ commute.
Thus all predictions from Eq.~(\ref{eq:vNstate}) are preserved in the equivalent state
\begin{equation}
\begin{array}{rcccl}
\{v(M')M',&M_2'\I,&\ldots,&M_n'\I,&v(M)M\I;\\
cd\I\ldots\I X,&C_2'\I,&\ldots,&C_n'\I,&cC_1'\I\}.
\end{array}
\end{equation}

Since $M'=\I\ldots \I Z$, the measured system and the pointer system are now completely independent, so the pointer system can be removed without affecting the measured system.
This is a direct consequence of the COM contextual measurement update when measuring $M'$.
Removing the pointer system and reordering the generators we obtain
\begin{equation}
\begin{array}{rccl}
\{v(M)M,&M_2',&\ldots,&M_n';\\
cC_1',&C_2',&\ldots,&C_n'\}.
\end{array}
\end{equation}
This is identical to Eq.~\eqref{eq:out}, except that the phase $c$ originates from the initial state of the pointer system in Eq.~\eqref{eq:in}. 
This concludes the generic proof that the state update of the measured system in the contextual Heisenberg microscope is not random, it is pre-determined by the pointer state, just as in the original Heisenberg microscope.\hfill $\square$

\bibliography{Mittbibliotek}

\end{document}